# Access Interfaces for Open Archival Information Systems based on the OAI-PMH and the OpenURL Framework for Context-Sensitive Services

Jeroen Bekaert[1,2], and Herbert Van de Sompel[1]

[1] Digital Library Research & Prototyping Team, Los Alamos National Laboratory,
MS P362, PO Box 1663, Los Alamos, NM  87544-7113, US
`{jbekaert, herbertv}@lanl.gov`
[2] Dept. of Architecture and Urbanism, Faculty of Engineering, Ghent University,
Jozef-Plateaustraat 22, 9000 Gent, Belgium
`{jeroen.bekaert}@ugent.be`

**Abstract.** In recent years, a variety of digital repository and archival systems have been developed and adopted. All of these systems aim at hosting a variety of compound digital assets and at providing tools for storing, managing and accessing those assets. This paper will focus on the definition of common and standardized access interfaces that could be deployed across such diverse digital respository and archival systems. The proposed interfaces are based on the two formal specifications that have recently emerged from the Digital Library community: The Open Archive Initiative Protocol for Metadata Harvesting (OAI-PMH) and the NISO OpenURL Framework for Context-Sensitive Services (OpenURL Standard). As will be described, the former allows for the retrieval of batches of XML-based representations of digital assets, while the latter facilitates the retrieval of disseminations of a specific digital asset or of one or more of its constituents. The core properties of the proposed interfaces are explained in terms of the Reference Model for an Open Archival Information System (OAIS).

## 1. Introduction

The Open Archival Information System (OAIS) Reference Model [1] addresses a wide range of information preservation functions, including ingest, storage, management, and access. The Reference Model also conceptually identifies the internal and external interfaces to these archival functions. Of specific importance for this paper is the OAIS Coordinate Access Activities function, which provides the conceptual interfaces to the holdings of an archive compliant with the OAIS Reference Model. Several pre-defined categories of access requests are distinguished, including Order requests aimed at returning Dissemination Information Packages (DIPs). Another type of request that is of particular interest in the context of this paper is a Dissemination Request. This type of request aims at returning a dissemination of one or more constituents of a digital asset. While an Order request is mainly intended for machine-based consumption, the response of a dissemination request is typically presented to an end user.

The concept of a common access interface to a repository of digital assets is also recognized in the Kahn/Wilensky framework [2]. There, the conceptual Repository Access Protocol is introduced to allow requesting disseminations of a 'digital object' by specifying its identifier, a service request type, and a set of additional parameters. The need for common access interfaces has also been expressed and explored by various projects. For example, the Networked European Deposit Library (NEDLIB), a project aimed at defining a workflow for ingesting, storing and accessing content in deposit systems for electronic publications, raised the need to explore possible '*standardized techniques for the content transfer from publishers to libraries, following the OAIS Reference Model*' [3]. The Joint Information Systems Committee (JISC) Digital Repository Programme concludes that '*a technical implementation of the OAIS DIP/SIP interfaces*' is one of many areas where additional standards or specifications are needed [4]. And, in a paper describing a project aimed at mirroring the collection of the American Physical Society (APS) at the Research Library of the Los Alamos National Laboratory, the authors observe the lack of standardization in repository access mechanisms that the Library experiences when uploading content from scholarly publishers. They introduce a standards-based (OAI-PMH) interface to the APS repository to allow recurrent and accurate transfer of digital assets, and suggest the approach could be deployed beyond the context of the described project [5].

Because the importance of common and standardized access interfaces to digital asset repositories is well recognized, it is somehow surprising to find that, so far, no cross-community solutions have been proposed and deployed in this realm. Indeed, typically, a different access interface exists per repository system. In this paper, in order to try and help alleviate this impasse, two standard-based repository access mechanisms are proposed that could be deployed across systems and communities. The proposed interfaces are based on the two formal specifications that have recently emerged from the Digital Library community: The Open Archive Initiative Protocol for Metadata Harvesting (OAI-PMH) [6] and the NISO OpenURL Framework for Context-Sensitive Services (OpenURL Standard) [7]. As will be described, the former allows for the retrieval of batches of XML-based representations of digital assets, while the latter facilitates the retrieval of disseminations of a specific digital asset or of one or more of its constituents.

To allow for a good understanding of the proposed interfaces, Section 2 of this paper describes some crucial concepts from the OAIS Reference Model, Section 3 maps those concepts to a few real life repository systems, and Section 4 describes the essence of the OAI-PMH and the NISO OpenURL Standard. Section 5 introduces the proposed interfaces, and Section 6 looks at the applicability of the proposed interfaces for the repository systems described in Section 3. The paper concludes by discussing the potential usability of the proposed interfaces and looking at possible areas of future work.

## 2. An OAIS perspective on the representation, identification and versioning of digital assets

The Reference Model for an Open Archival Information System (OAIS) [1] developed by the CCSDS has become a foundation for thinking about problems in the digital preservation domain. The OAIS Reference Model defines both a Functional Model and an Information Model. The Functional Model outlines the range of functions that need to be undertaken by a compliant archive, such as access, archival storage, and ingest. The Information Model defines broad types of information that are required in order to preserve and access the information stored in an archive. Both the Functional Model and the Information Model define useful abstract concepts but not a blueprint for an implementation of an archival system. The organizational and technical choices of how to implement the abstract OAIS concepts in actual concrete environments is left to the communities involved.

Although the focus of the proposed access interfaces is not only archives tasked with the long-term preservation of digital information, but rather repository systems in general, the Reference Model for an OAIS provides well-defined terminology that allows expressing a variety of properties of systems that store digital assets. Therefore, in this paper, terms of the Informational and Functional Models will be used to describe the core characteristics of the proposed interfaces. A description of each of those terms is provided below; some are also depicted in Figure 1.

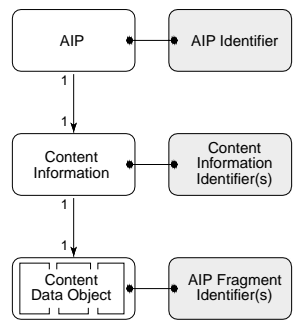

**Fig. 1.** Archival Information Package, Content Information, and Content Data Object

*Content Information* is information that is the original target of preservation in an OAIS environment. In the information science domain, Content Information is typically referred to as a digital asset or a digital object; it is the information object of primary interest to an end user.

Content Information is the binding of a *Content Data Object* with *Representation Information* that supports the actual representation of the Content Data Object. A Content Data Object is a sequence of bits that is typically implemented as one or more *files*. It is fair to state that these Content Data Object files can be considered the actual constituents of the Content Information.

An *Information Package* is a container that binds the Content Information with associated *Preservation Description Information*. Preservation Description Information is information that is essential to adequately preserve the particular Content Information to which it is bound. An Information Package is serialized using *Packaging Information*. This Packaging Information also provides local hooks (often called 'Fragment Identifiers') into the Information Package to allow accessing each file of which the Content Data Object of the Content Information consists.

Content Information can have one or more identifiers, each of which is named a *Content Information Identifier*; these identifiers reside under the *Reference Information* sub-category of the Preservation Description Information.

An Information Package has a single *Information Package Identifier*. The value of this identifier must be unique within a (federation of) archive(s). In contrast with the Content Information Identifier, the Information Package Identifier is an internal property of the repository, and hence, is typically not exposed to downstream applications or end-users.

The Information Model for an OAIS recognizes three subtypes of the Information Package: the *Archival Information Package* (AIP), the *Submission Information Package* (SIP), and the *Dissemination Information Package* (DIP). The definitions of these types of Information Packages are based on the function of the archival process that uses the Information Package, and on the translation from one Information Package to another as it passes through the archival process. A distinction is made between Information Packages that are submitted to an archive (i.e. SIP), Information Packages that are subsequently stored and preserved by an archive (i.e. AIP), and those that are disseminated from an archive (i.e. DIP). In order for archives in a federation to be able to exchange Information Packages they must support at least one common *DIP/SIP format*.

According to the OAIS Functional Model, whenever the Content Information or its Preservation Description Information is updated, a new AIP must be created. Dependent on the nature of the update, a distinction is made between several types of AIPs. Of special interest for this paper are the notions of Version and Edition. A *Version* is an AIP that results from applying a transformation, induced by a preservation strategy, on the Content Information of a source AIP. An *Edition* is an AIP that results from increasing or improving the Content Information of a source AIP; for example by removing a few typographic errors from one of the constituents of the Content Information. Both an AIP Version and an AIP Edition are candidates to replace the source AIP from which they are derived. No matter which type of update the Content Information or the Preservation Description Information of a source AIP undergoes, the result is a new AIP that receives a new, unique AIP Identifier. Note that in both cases, the Preservation Description Information needs to be updated to provide information about the source AIP, and to describe what was done and why. The Content Information Identifier may remain untouched.

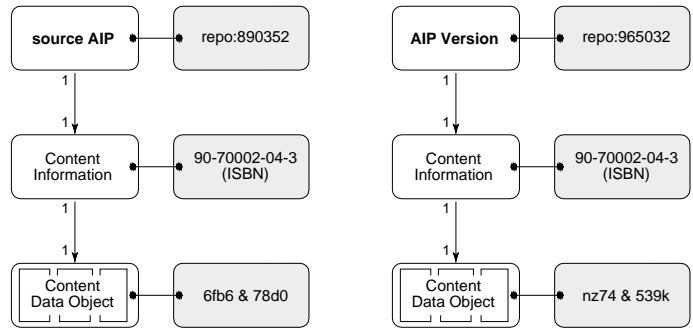

**Fig. 2.** Archival Information Package Versions

Because it may be of historical interest, especially in the context of digital preservation, to retain previous Versions or Editions of AIPs, it results that, within a single archive, several versions of specific Content Information may exist. All these versions share a Content Information Identifier, yet have a different Archival Package Information (AIP) Identifier. Each version of specific Content Information can be retrieved using the AIP Identifier of the AIP that encapsulates the particular version. An example of a source AIP and an AIP Version is depicted in Figure 2. As can be seen, both AIPs share the same Content Information Identifier (i.e. the ISBN number 90-70002-04-3); yet each AIP is identified using a different AIP Identifier (i.e. the identifiers repo:890352 and repo:965032, respectively).

## 3. The representation, identification and versioning of digital assets in real-life repositories

In recent years, a variety of digital repository and archival systems have been developed and adopted. Of specific interest in the current technological environment are systems that are capable of hosting compound digital assets, consisting of one or more datastreams of a variety of MIME media types. Examples of such information systems that will be considered here are *aDORe*, a repository architecture designed and implemented at the Research Library of the Los Alamos National Laboratory [8], *DSpace*, a digital repository system jointly developed by MIT Libraries and Hewlett-Packard (HP) [9,10], and *Fedora*, an open-source software jointly developed by Cornell University and the University of Virginia [11,12]. All of these systems aim at hosting a variety of compound digital assets and at providing tools for storing, managing and accessing those assets. In spite of their similar goals, each of these systems comes with its own perspective on how to accomplish them.

This section explores how specific properties of the data models that underly aDORe, DSpace and Fedora map to the concepts of the OAIS Information Model discussed in Section 2. The focus hereby is on those properties that are core to the proposed access interfaces, namely on the handling of identifiers and versions of digital assets. Figure 3 summarizes the findings described hereafter.

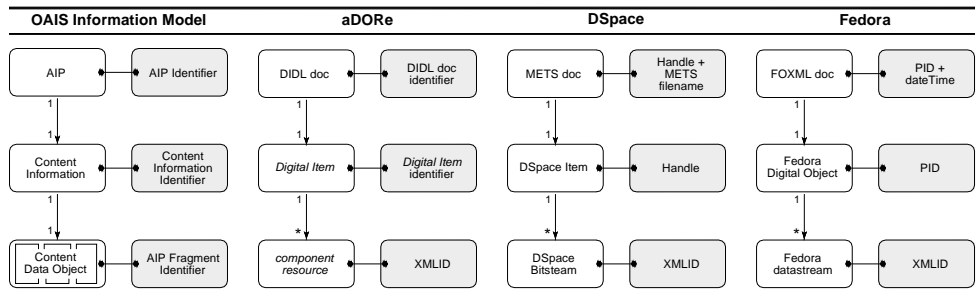

**Fig. 3.** Mapping concepts from the OAIS Information Model to the aDORe, DSpace, and Fedora systems

- ***aDORe*** [8]: Compound digital assets stored in the aDORe environment are represented according to the MPEG-21 DID Abstract Model and serialized using the MPEG-21 DIDL syntax. In MPEG-21, the digital asset itself is called a *Digital Item* and is considered the primary information of interest to an end user in the MPEG-21 environment. The XML document that packages and serializes the *Digital Item* is referred to as a DIDL document [13,14,15]. As such the aDORe *Digital Items* map to OAIS Content Information; the DIDL XML documents that package the *Digital Items* map to OAIS Archival Information Packages. The MPEG-21 DIDL syntax is the Packaging Information.

    aDORe has two parallel identification mechanisms. The first identification mechanism pertains to *Digital Items*. This type of identifier is typically associated with a digital asset at the moment of its creation by a publisher, and, hence already exists when the asset is ingested into aDORe. If it does not yet exist, it is created upon receipt. Clearly, this type of identifier maps to the concept of OAIS Content Information Identifiers. A second identification mechanism pertains to the DIDL XML documents that package the *Digital Items*. These identifiers are minted during ingestion into aDORe. They are unique within the aDORe repository, and even globally unique through the use of the info URI scheme [16]. This type of identifiers is mainly used for repository management purposes; it directly maps to the concept of OAIS AIP Identifiers.

    During creation of a DIDL XML document, each constituent datastream of a *Digital Item* is conveyed as an MPEG-21 DID *component/resource* constructs, and is

accorded a Fragment Identifier. As a result, each constituent datastream becomes addressable in the aDORe repository using a combination of the AIP Identifier of the DIDL document in which it is contained, and its own Fragment Identifier.

In the aDORe repository, whenever a new version of a previously ingested digital asset is ingested, a new DIDL XML document is created for it; existing DIDL documents are never updated or edited. As such, a version of a digital asset (aka *Digital Item*) can be directly retrieved using the identifier of the DIDL document that packages the transformed or modified content. All these versions of a specific digital asset share the same *Digital Item* identifier. This approach closely resembles the versioning concepts defined by the OAIS Functional Model.

- *DSpace* [9,10]: Compound digital assets stored in the DSpace repository are organized using the DSpace Data Model. A digital asset is typically represented as a DSpace Item; datastreams aggregated by the digital asset are called DSpace Bitstreams. The current DSpace digital repository system (release 1.3.1 – August 2005) instantiates the DSpace Data Model using linked tables in a relational database. DSpace Bitstreams are stored in a file system. Every DSpace Item receives a persistent unique identifier; DSpace uses the Handle System for minting, managing and resolving these identifiers. It is important to note that DSpace treats the identifiers that were assigned to digital assets before their ingestion into DSpace (i.e. URL, DOI, ISBN, etc.) as descriptive metadata (`DC:identifier`), not as identifiers that can easily be mapped to identifier concepts in the OAIS Information Model. This is in contrast with aDORe that treats these identifiers as the OAIS Content Information Identifiers. Also, the current release of the DSpace system does not seem to provide an unambiguous solution for the versioning of stored digital assets. While it is argued that a separate DSpace Item could be created for each distinct version of a digital asset [17], this approach does not allow for multiple versions of a digital asset to share the same identifier. Overall, it seems difficult to unambiguously map the manner in which the current version of DSpace handles identifiers and versions to corresponding concepts of the OAIS Reference Model.

    However, at the time of writing, plans exist to create a new version of the DSpace repository system (release 2.0) [18]. Our discussion of the mapping of DSpace concepts to those of the OAIS Reference Model will focus on this planned version. In the planned version, the DSpace Data Model would remain untouched, but it would introduce a flat file storage mechanism in which each DSpace Item is represented as an XML document conformant with the Metadata Encoding and Transmission Standard (METS) syntax [19]. The main reason for replacing the current relational database structure with the METS-based solution is to ease various preservation related tasks, including disaster recovery, versioning control, and data replication. In this revised approach, according to the OAIS Information Model, the DSpace Items that represent the actual content can be considered Content Information. A METS XML document that represents and serializes the content as a storable package can be considered the Archival Information Package. The METS syntax is the Packaging

Information. In the planned release, different versions of a stored digital asset would be represented by different METS documents, one per version. Following this approach, several versions of a DSpace Item may share the same handle identifier; they will be distinguished by a different METS document. From this discussion, it follows that in the planned version 2.0 of the DSpace system, the handle that is assigned to each DSpace Item maps to the concept of the OAIS Content Identifier. A unique METS file name or a unique identifier for a METS file would map to concept of an OAIS AIP Identifier. The exact details with this respect remain a topic of further study by the DSpace group [18].

- *Fedora* [11,12]: Compound digital assets stored in the Fedora repository (release 2.0) are represented according to the Fedora Digital Object Model [20] and encoded using the FOXML syntax [21]. The digital asset itself is referred to as a Fedora Digital Object; it maps to the OAIS concept of Content Information. The serialization of a Fedora Digital Object in FOXML is called a FOXML document. This serialization maps to the OAIS concept of Archival Information Package. The FOXML syntax can be considered the Packaging Information.

    The Fedora system does not make an explicit distinction between identifiers accorded to the actual content (i.e. Content Information Identifiers), and identifiers pertaining to stored packagings of the content (i.e. Information Package Identifiers). Both the Fedora Digital Object and its representation in FOXML share the same unique persistent identifier, referred to as the 'PID' (Persistent Identifier). PIDs may be minted by a Fedora repository or may be user-defined; the latter allows for the use of identifiers that were assigned to digital assets prior to their ingestion into Fedora. During the creation of a FOXML document, each datastream constituting a Fedora Digital Object is accorded an XML Fragment Identifier. As a result, these constituent datastreams become addressable in the Fedora repository using a combination of the PID and their own Fragment Identifier.

    In the Fedora repository, a modification made to a constituent of a Fedora Digital Object results in the creation of a new version of that constituent. Fedora does not create a new FOXML document when a new version of a constituent of a Digtal Object becomes available. Instead, each specific constituent is versioned in the source FOXML document through the assignment of a local key that conveys a dateTime of creation or update. In this versioning approach, the PID of the Fedora Digital Object remains constant [22]. Because multiple versions of a Fedora Digital Object may share the same PID, and because a PID is considered the primary identification mechanism of a Fedora Digital Object for downstream application, the PID maps to the OAIS concept of Content Information Identifier. In addition, a version of a Fedora Digital Object can be uniquely identified using the combination of a PID and a specific dateTime key. This combination maps to the OAIS concept of the AIP Identifier.

    For completeness, it is worthwhile mentioning that from the above discussion it follows that in terms of the OAIS, a FOXML document in Fedora is really a collection

of AIPs in which each AIP holds a version of a Fedora Digital Object. However, this refined perspective does not change the aforementioned reasoning regarding identifiers.

## 4. The OAI-PMH and the OpenURL Framework for Context-Sensitive Services

In this paper, the potential of two existing specifications will be explored to define common access interfaces to information systems: The Open Archives Protocol for Metadata Harvesting (OAI-PMH), a widely adopted specification that allows for the selective harvesting of metadata, and the OpenURL Framework for Context Sensitive Services (NISO OpenURL), a recent NISO Standard, formally known as ANSI/NISO Z39.88-2004. Both are briefly described below.

### 4.1 The Open Archival Initiative Protocol for Metadata harvesting (OAI-PMH)

The OAI-PMH [6] is a protocol that allows for the recurrent harvest of XML-based metadata from one place to another. An OAI-PMH repository exposes a collection of metadata records. A harvester issues OAI-PMH protocol requests, in order to harvest XML metadata. The OAI-PMH builds on existing standards, notably the IETF Hypertext Transfer Protocol (HTTP), and the W3C Extensible Markup Language (XML) syntax for encoding the exchanged metadata. OAI-PMH harvesters may request information from OAI-PMH repositories using a standard set of six OAI-PMH verbs. OAI-PMH requests are transmitted according the rules of HTTP 1.0, with requests specified using URL-encoded parameters and responses delivered in strictly validifiable XML.

The OAI-PMH 2.0 solution is based on a data model – depicted in Figure 4 – that helps specifying the semantics of the six protocol requests. In what follows, OAI-PMH entities of the data model are written in *italic* font, while OAI-PMH protocol requests are written in `courier`:

- At the very top is a digital *resource* about which an OAI-PMH repository exposes metadata. By definition, *resources* themselves are outside of the scope of the OAI-PMH.
- Listed below the *resource* is the *item*. The *item* is the highest-level entity within the scope of the OAI-PMH. In essence, the *item* is the entry point to all available *metadata* pertaining to a *resource*. In the protocol, the *item* is uniquely identified by an OAI-PMH *identifier*. All possible *metadata records* available from a single *item* share the same OAI-PMH *identifier*.

- Below the *item*, several *records* are shown. A *record* is *metadata* in a specific *metadata format*. A specific *record* in the OAI-PMH is unambiguously identified by means of the combination of the OAI-PMH *identifier* (of the *item*), the *metadata format* used for the dissemination of the *metadata*, and the OAI-PMH *datestamp* of the *metadata*. The *datestamp* is the date and time of creation or modification of the *metadata*. Note that the *datestamp* is a property of the *metadata* record, not of the *item* as used to be the case in previous protocol versions. This reflects the fact that *metadata* of various *metadata formats* may be made available and may be modified independently, thus having different *datestamps*.
- The OAI-PMH also defines a *set* as an optional construct for grouping *items* for the purpose of selective harvesting. Repositories may organize *items* into *sets*. A *set* organization may be flat, i.e. a simple list, or hierarchical. Multiple, parallel, *set* structures may exist.

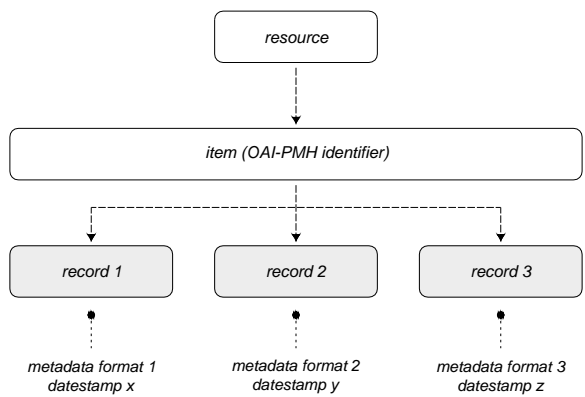

**Fig. 4.** The OAI-PMH data model

The OAI-PMH consists of six verbs, three of which reveal the characteristics of the repository (`ListMetadataFormats`, `ListSets`, and `Identify`) and three verbs for extracting metadata from the repository (`GetRecord`, `ListRecords`, `ListIdentifiers`). Each OAI-PMH verb requires and/or allows the use of certain parameters that further define the exact nature and details of the request. The OAI-PMH defines three supporting protocol requests that are aimed at helping a harvester understand the nature of an OAI-PMH repository:

- `Identify`: This verb is used to retrieve information about an OAI-PMH repository; an important information element returned in the response to the `Identify` request is the granularity of the *datestamp* supported by the repository (day-level or seconds-level).

- `ListMetadataFormats`: This verb is used to retrieve the *metadata formats* available from a repository.
- `ListSets`: This verb is used to retrieve the *set* structure of a repository. This information is useful for selective harvesting.

The OAI-PMH defines three further protocol requests that are aimed at the actual harvesting of XML-based metadata:
- `ListRecords`: This verb is used to harvest *records* from a repository. Optional arguments permit selective harvesting of *records* based on *set* membership and/or *datestamp*.
- `GetRecord`: This verb is used to retrieve an individual *record* from a repository. The verb has two required arguments; one argument specifies the OAI-PMH *identifier* of the *item* from which the *record* is requested; the other conveys the *metadata format* of the *metadata* that should be included in the *record*.
- `ListIdentifiers`: This verb is an abbreviated form of `ListRecords`, retrieving only *identifiers*, *datestamps* and *set* information.

Data providers process the OAI-PMH requests and reply with appropriate OAI-PMH responses, which are always in the form of valid XML conforming to top-level XML schemas defined by the OAI-PMH.

Due to its origins in the realm of resource discovery, the OAI-PMH mandates the support of the Dublin Core metadata format, but strongly encourages supporting more expressive formats. As a result, any *metadata format* can be used as long as it is defined by means of an XML Schema. In typical use cases, the exposed *metadata* is descriptive, and is expressed by means of *metadata formats* of varying complexity, such as simple Dublin Core, or MARCXML. However, recently, new use cases have emerged that reveal a more liberal interpretation of what constitues *metadata*. For example, at the Research Library of the Los Alamos National Laboratory, various projects have been carried out that explore the use of XML-based complex objects formats in combination with the OAI-PMH. Examples include the use of OAI-PMH to export DIDL document from the aDORe repository [8], the use of OAI-PMH to transfer digital assets from the American Physical Society to the Los Alamos National Laboratory [5], and the use of the OAI-PMH to expose content accessible from Apache Web servers [23].

The *metadata formats* supported by the OAI-PMH Interface proposed in this paper are DIP formats. The *metadata records* returned in response to OAI-PMH requests are DIPs derived from stored AIPs. Each DIP is unambiguously identified by means of the combination of an OAIS DIP format, an OAI-PMH *identifier,* and the OAI-PMH *datetstamp* of the OAIS DIP. The OAI-PMH *identifier* in the proposed Interface can be expressed in terms of OAIS AIP Identifiers or OAIS Content Information Identifiers.

**4.2 The NISO OpenURL Framework for Context-Sensitive Services**

The NISO OpenURL Framework Standard [7] defines an architecture for creating *OpenURL Applications*. An *OpenURL Application* is a networked service environment, in which packages of information are transported over a network. The main purpose of the transportation of these packages is to request and obtain context-sensitive services pertaining to a referenced resource. In order to do so, each package describes the referenced resource itself, the network context in which the resource is referenced, and the context in which the service request takes place.

The NISO OpenURL Framework Standard originated in the scholarly information community. Within that community, the initial OpenURL 0.1 specification [24] – the precursor of the NISO OpenURL standard – was introduced for the specific purpose of reference-linking (i.e. referencing an article citation to the full text of the article) and was targeted at facilitating the provision of context-sensitive service links for popular types of scholarly works such as journal articles and books. Hereby, identifiers and metadata describing the work are conveyed using a controlled-vocabulary HTTP GET request to a user-specific linking server, which uses a rules-based approach to provide an agent or end user with appropriate services pertaining to the work.

A generalization of the essential components of the initial OpenURL 0.1 solution, beyond the scholarly information environment, inspired the very nature of the NISO OpenURL standard [7]. The NISO OpenURL Framework Standard allows for expressing requests for the delivery of context-sensitive services pertaining to whichever type of resource referenced in a networked environment. Again, the main pre-requisite for this extension is the existence of identifiers and/or metadata that describe the referenced resources and its network context.

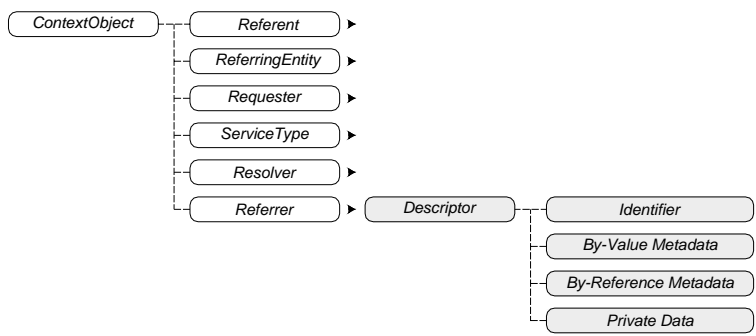

**Fig. 5.** Each *Entity* of a *ContextObject* is specified using a *Descriptor*; four *Descriptor* types can be used simultaneously.

To that end, the NISO OpenURL Standard introduces the notion of a *ContextObject*; an abstract information construct that contains descriptions of various *Entities* involved in the process of requesting context-sensitive services. The different *Entities* are shown in Table 1 and are also depicted in Figure 5. In what follows, pre-defined terms provided by the NISO OpenURL standard are written in *italic* font.

| *Entity* | Definition |
|---|---|
| A *Referent* | The *Entity* that is referenced in a networked environment and about which the *ContextObject* is created |
| A *ReferringEntity* | The *Entity* that references the *Referent* |
| A *Requester* | The *Entity* that requests services pertaining to the *Referent* |
| A *ServiceType* | The *Entity* that defines the type of service requested |
| A *Resolver* | The *Entity* at which a request for services is targeted |
| A *Referrer* | The *Entity* that generates the *ContextObject* |

**Table 1**. The NISO OpenURL *Entities* of a *ContextObject*

A *ContextObject* can be transported to a networked system, named a *Resolver*, in order to request services (expresssed by a *ServiceType*) pertaining to the *Referent* described in it. To decide upon the nature of such services, the *Resolver* may take *Entities* other than the *Referent* and *ServiceType* into account. These other *Entities* are *ReferringEntity*, *Requester*, and *Referrer*. Each *Entity* of the *ContextObject* can be described by means of so-called *Descriptors*. As depicted in Figure 5 and listed in Table 2, the Standard distinguishes between *Identifier Descriptors*, *Metadata Descriptors* and *Private Data Descriptors.*

| **Descriptor type** | **Definition** |
|---|---|
| An *Identifier* | This *Descriptor* unambiguously specifies the *Entity* by means of a Uniform Resource Identifier (URI). This URI either points to the *Entity* itself or to metadata that specify the *Entity*. |
| *By-Value Metadata* | This *Descriptor* specifies properties of the *Entity* by the combination of: 1) a URI reference to a *Metadata Format*; and 2) a particular instance of metadata about the *Entity* expressed according to this *Metadata Format*. |
| *By-Reference Metadata* | This *Descriptor* specifies properties of the *Entity* by the combination of: 1) a URI reference to a *Metadata Format*; and 2) the network location of a particular instance of metadata about the *Entity* expressed according to this *Metadata Format*. |
| *Private Data* | This *Descriptor* specifies information about the *Entity* using out-of-band technology. The *Resolver* and the *Referrer* have a common understanding of the *Descriptor* based on a bilateral agreement. |

**Table 2**. The NISO OpenURL *Descriptor* types

The NISO OpenURL standard makes a clear distinction between the abstract definition of the above concepts, their concrete representation, and the protocol by which such representations are transported. The OpenURL Framework allows for a

*ContextObject* to be represented in many different *Formats* and currently, a Key/Encoded-Value (KEV) representation and an XML representation have been defined. A representation of a *ContextObject* is transported to a *Resolver*, in order to request services pertaining to the *Referent* described in it. The transport of a *ContextObject* can occur using various network protocols, and currently transport over HTTP and HTTPS have been defined.

To address the issue of open-endedness, and allow for the creation of highly interoperable solutions, an *OpenURL Framework Registry* is introduced [http://www.openurl.info/registry], providing a mechanism for the public disclosure of specific selections for the representation and transportation of *ContextObjects*. In general, a community or application domain defines an *OpenURL Application* by constraining the type and numbers of *Entities* and *Descriptors* pertaining to a *ContextObject* and by selecting entries from the *Registry* to represent and transport the *ContextObjects*. If necessary and/or desired, the community may define and register new entries.

## 5. An OAIS perspective on the access to digital assets

In this paper, two access interfaces for an OAIS are proposed that could be deployed across systems and communities. One interface is based on the OAI-PMH, the other on the NISO OpenURL Framework Standard. These interfaces support the following types of repository access:

- Requests aimed at returning OAIS Dissemination Information Packages. In the OAIS Reference Model, this type of request is referred to as an **Order**. An Information Package Order identifies one or more OAIS Archival Information Packages (AIPs) of interest, and specifies how these OAIS AIPs are to be mapped into OAIS Dissemination Information Packages (DIPs). In response to an Order, an OAIS compliant archive provides all or a part of an OAIS AIP to a consuming archive in the form of an OAIS DIP. Of specific interest, in particular in the context of digital preservation, is the retrieval of individual versions of stored OAIS Content Information.
- Requests aimed at returning a dissemination of one or more (parts of an) OAIS Content Data Object(s). The response to this request is a MIME-typed bitstream. Again, the nature of a dissemination may vary dependent on the version of the OAIS Content Information (and hence of its OAIS Content Data Object). Because the OAIS Reference Model does not introduce a specific term to refer to this type of request, we will henceforth refer to it as a **Content Data Object Dissemination Request,** or **Dissemination Request** in short.

As described in Section 2 of this paper, the OAIS Information Model recognizes the existence of two parallel identification mechanisms to address information stored in an archive or repository. One mechanism is directly related to the identification of OAIS Content Information using the OAIS Content Information Identifier; the other mechanism uses the OAIS AIP Identifier to identify an OAIS AIP stored in an archival system. While the same OAIS Content Information Identifier may be shared by multiple sets (or versions) of OAIS Content Information, the OAIS AIP Identifier is considered unique within an information system. It follows that access interfaces to an OAIS compliant system could be centered around both identification mechanisms. This paper will focus on the definition of access interfaces based on OAIS Content Information Identifiers only. A first interface is based on the OAI-PMH, and allows for the Order of batches of DIPs. A second interface is based on the NISO OpenURL Framework. For this interface, two levels of conformance are defined. The first level of conformance allows to Order individual OAIS DIPs from an information system; the second level is directly related to Dissemination Requests for (parts of) a Content Data Object.

### 5.1 Interface #1: Ordering OAIS DIPs using OAI-PMH

A first Interface (henceforth referred to as *Interface #1*) allows for the retrieval of OAIS DIPs from an information system using the OAI-PMH protocol. The identifier of the OAIS Content Information that is packaged by an OAIS DIP serves as the OAI-PMH *identifier*. The response returned by the interface is an OAI-PMH *record*. Each *record* physically embeds the requested OAIS DIP as OAI-PMH *metadata*. Following the OAI-PMH specification, the OAIS DIP must be delivered in strictly valid XML. The specifics of this interface are described below.

The OAI-PMH Interface #1 of the OAIS that hosts the OAIS AIPs has the following characteristics:

- The baseURL of the OAI-PMH interface is the HTTP address `baseURL(OAIPMH_CIID)`.
- The OAI-PMH *identifiers* used by Interface #1 are the OAIS Content Information Identifiers of the OAIS Content Information packaged by the OAIS AIPs stored in the repository.
- The OAI-PMH *datestamps* used by the OAI-PMH interface are the datetime of creation of the OAIS AIPs. Note that, because the OAIS Reference Model requires the creation of a new AIP (instead of an update of an OAIS AIP) for every new version of OAIS Content Information, the OAI-PMH datestamp of a given AIP will never change once the OAIS AIP has been created. Because of the nature of the OAI-PMH, a request for an OAIS DIP containing OAIS Content Information with a specific OAIS Content Information Identifier will result in an OAIS DIP derived from the most recent OAIS AIP that packages the identified OAIS Content Information; that is from the most recent version of the OAIS Content Information.

- The natively supported *metadata format* is an XML-based OAIS DIP format. Potentially, multiple OAIS DIP formats could be supported by Interface #1. Various XML-based Packaging formats have emerged over the last several years, some of which have been standardized. Examples include, the ISO MPEG-21 Digital Item Declaration Language (MPEG-21 DIDL) [13,14,15], the Metadata Encoding and Transmission Standard (METS) [19], the IMS Content Packaging XML Binding [26], and the XML Formatted Data Units (XFDU), a pre-standard developed by CCSDS Panel 2 [25].
- The supported granularity of Interface #1 is seconds-level.
- *Set* structures may be supported for grouping OAIS AIPs for the purpose of selective harvesting.

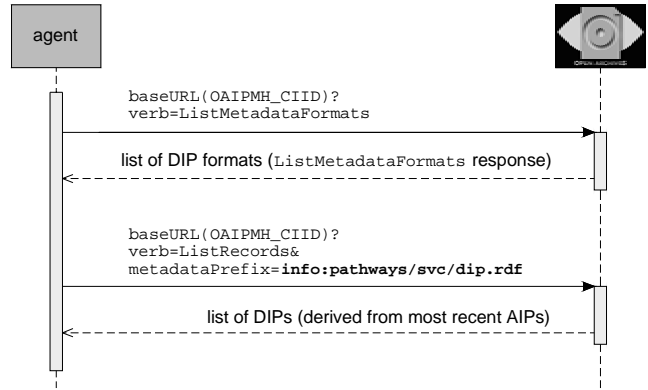

**Fig. 6.** Sequence Diagram of Interface #1:
Ordering OAIS DIPs using OAIS Content Information Identifiers and OAI-PMH

The interaction of a downstream OAI-PMH harvester with the repository through the OAI-PMH interface is illustrated in Figure 6 and explained below. Typically, the interaction is a two-step approach:
- Firstly, the OAI-PMH harvester issues a `ListMetadataFormats` request against the OAI-PMH Interface #1 of the OAIS. In response, the OAI-PMH harvester receives a list of supported XML-based OAIS DIP formats. The request looks as follows:

```
[BaseURL(OAIPMH_CIID)?
verb=ListMetadataFormats]
```

- Once a list of OAIS DIP formats has been obtained, OAI-PMH harvesters can retrieve OAIS DIPs from the OAI-PMH interface of the information system by issuing an OAI-PMH `GetRecord` or `ListRecords` request. Each OAIS DIP is provided as an OAI-PMH *record*. The *metadata format* of the OAIS DIP must correspond with one of the OAIS DIP formats retrieved from the repository by issuing the `ListMetadataFormats` request. Each *record* is uniquely identified by the combination of an OAIS Content Information Identifier of the OAIS Content Information packaged by the OAIS AIP (and from which the OAIS DIP will be derived) as the OAI-PMH *identifier*, an OAIS DIP format as the OAI-PMH *metadata format*, and the datetime at which the OAIS AIP containing the OAIS Content Information (and from which the OAIS DIP will be derived) has been created. It is very important to note that, as has been explained in Section 2 of this paper, multiple OAIS AIPs may exist that package a set of OAIS Content Information sharing the same OAIS Content Information Identifier. Because of the nature of the OAI-PMH, a request for an OAIS DIP that packages OAIS Content Information with a specific OAIS Content Information Identifier will result in an OAIS DIP derived from the most recent OAIS AIP that packages the identified OAIS Content Information.

An example of both a `GetRecord` and a `ListRecords` request is shown below. A `GetRecord` request automatically results in the retrieval of a single OAIS DIP that is derived from the most recent OAIS AIP that packages a set of OAIS Content Information with OAIS Content Information Identifier `ContentInfoIdentifier`. The `ListRecords` request results in a list of OAIS DIPs, in which each OAIS DIP is derived from the most recent OAIS AIP that, given a specific OAIS Content Information Identifier, has been created within the bounds of the (optional) `from` and `until` arguments. `info:pathways/dip.rdf` identifies an OAIS DIP format that is available from the OAI-PMH Interface #1.

```
[BaseURL(OAIPMH_CIID)?
verb=GetRecord&
identifier=ContentInfoIdentifier&
metadataPrefix=info:pathways/svc/dip.rdf]
```

and

```
[BaseURL(OAIPMH_CIID)?
verb=ListRecords&
from=T1&until=T2&
metadataPrefix=info:pathways/svc/dip.rdf]
```

For reasons of completeness, it should be noted that the proposed interface equates the identifier of the OAI-PMH *item* with the identifier of the OAI-PMH *resource*. While the *item* and *resource* are two different entities in the OAI-PMH data model, the OAI-PMH does not specify the nature of the OAI-PMH *resource* nor of its identifier. Indeed, the nature of the *resource* and its identifier is outside the scope of the OAI-PMH specification and hence, may vary dependent on the application domain. In the context of this application, the identifier of the OAI-PMH *item* is set to match that of the OAI-PMH *resource*.

The power of Interface #1 lies in the possibilities it offers for downstream applications to harvest batches of OAIS DIPs that package the most recent sets of OAIS Content Information available in an information system and to keep the retrieved content up to date using a *datestamp* based harvesting strategy. Indeed, newly added and updated content can be harvested using a `ListRecords` request, by setting the value of the `from` parameter to the datetime of the last harvest that was conducted. It should be noted that Interface #1 does not facilitate the harvesting of all versions of OAIS Content Information, but rather only the most recent version. In order to support harvesting all versions, an interface similar to Interface #1, but based on OAIS AIP Identifier can be defined. This is, however, outside of the scope of this paper. For now, it suffices to mentioned that such an interface is being used in the aDORe repository work [8].

### 5.2 Interface #2: Ordering OAIS DIPs and requesting Disseminations using NISO OpenURL

A second interface (henceforth referred to as *Interface #2*) is compliant with the NISO OpenURL Framework standard. As described in Section 4.2, the NISO OpenURL Framework allows a community or application domain to define *OpenURL Applications* by formally describing the restrictions the implementation of the abstract *ContextObject* data structure, and on the choice of a mechanism to transport concrete *ContextObjects*. In this Section, we propose an *OpenURL Application* that allows to request disseminations from data stored in an information system, using OAIS Content Information Identifiers as the primary key in the request. A distinction is made between two levels of conformance.

- *Conformance Level 1*: An OpenURL *Resolver* compliant with Conformance Level 1 of this *OpenURL Application* allows for ordering individual OAIS DIPs from an information system. The *Referent* in this *OpenURL Application* is Content Information, and it is being described by means of an *Identifier Descriptor,* which is the OAIS Content Information Identifier of the OAIS Content Information. The response returned by this OpenURL *Resolver* is an OAIS DIP. A detailed description is provided in Section 5.2.1 and illustrated in Figure 7.
- *Conformance Level 2*: An OpenURL *Resolver* compliant with Conformance Level 2 of the proposed *OpenURL Application* allows requesting disseminations of datastreams (aka Content Data Object files). Again, the OAIS Content Information Identifier is used as the value of the *Referent Identifier Descriptor*. The response returned by this OpenURL *Resolver* is a MIME-typed stream. A detailed description is provided in Section 5.2.2 and depicted in Figure 8.

In the *OpenURL Application* described in this paper, *ContextObjects* are represented using the Key/Encoded-Value (KEV) *ContextObject Format* and are transported to an OpenURL *Resolver* using the HTTP(S) GET mode of the *By-Value OpenURL Transport*. It should be noted however that, because the OpenURL Standard is specified in a generic manner, the same interface can be implemented in many different ways as technologies evolve. For example, the concepts underlying the proposed Interface #2 can also be instantiated in an *OpenURL Application* in which *ContextObjects* are represented by means of the XML *ContextObject Format* and transported using an XML-based protocol such as SOAP. In essence, this means that the conceptual interface that underlies the proposed, concrete Interface #2 remains persistent over time.

### 5.2.1 Conformance Level 1: Ordering OAIS DIPs using NISO OpenURL

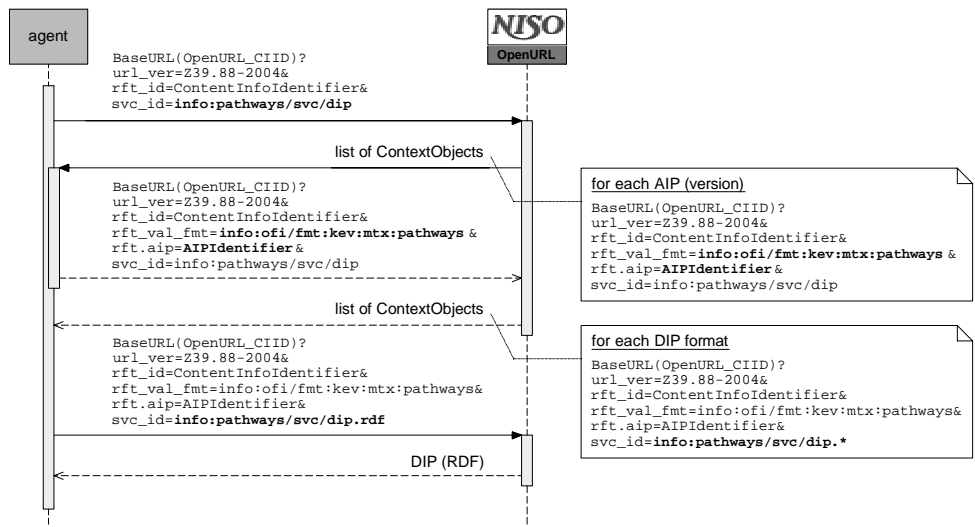

**Fig. 7**. Sequence Diagram of Interface #2, Conformance Level 1:
Ordering OAIS DIPs using OAIS Content Information Identifiers and NISO OpenURL

The Interface of an OpenURL *Resolver* compliant with Conformance Level 1 of the proposed *OpenURL Application* accepts two types of service requests. Both types of requests are expressed by means of a *ContextObject* that is transported towards the OpenURL *Resolver* at baseURL `OpenURL_CIID`:

- ***The interoperable OAIS DIP bootstrap service request***: The OpenURL *Resolver* of an information system compliant with Conformance Level 1 of this *OpenURL Application* must support the 'OAIS DIP bootstrap' request. This request is conveyed as a *ContextObject* with the following charachteristics:
  - The *Referent* of the *ContextObject* is OAIS Content Information stored (as an OAIS AIP) in the information system. The *Referent* is described by means of an

*Identifier Descriptor*. Its value is the OAIS Content Information Identifier of the OAIS Content Information in question.
- The *ServiceType* of the *ContextObject* is a service requesting a list of all OAIS DIP formats that can be provided for the *Referent*. The *ServiceType* is described by means of an *Identifier Descriptor* with the fixed value 'info:pathways/svc/dip'.
- The *ContextObject* may contain *Entities* other than *Referent* and *ServiceType*. These *Entities* offer the potential for describing, for example, properties of the agent that issues the request. This would allow tailoring the response to those properties.

As described in Section 2 of this paper, given a single OAIS Content Information Identifier, multiple OAIS AIPs may exist. As such, the first task of this *OpenURL Application*, in response to the initial 'OAIS DIP bootstrap' request, is to disambiguate between the various OAIS AIPs available from the information system for a given Content Information Identifier. Therefore, a separate process is started in which the *OpenURL Application* returns a list of all OAIS AIPs that can be provided for a given OAIS Content Information Identifier to the agent. This list is expressed as an XML container of *ContextObjects*; the syntax of the list must be valid against the XML Schema for the XML *ContextObject* Format [http://www.openurl.info/registry/docs/xsd/info:ofi/fmt:xml:xsd:ctx]. For each OAIS AIP (that packages OAIS Content Information identified by the given OAIS Content Information Identifier), a new *ContextObject* is provided. Each such *ContextObject* has the following charactersitics:
- The *Referent* of the *ContextObject* is an OAIS AIP containing the OAIS Content Information for which the initial OAIS DIP bootstrap service has been requested. The *Referent* is described by the combination of an *Identifier Descriptor* and a *By-Value Metadata Descriptor*. The value of the former is the OAIS Content Information Identifier as conveyed by the initial OAIS DIP bootstrap service. The latter conveys the OAIS AIP Identifier of the *Referent*. The *Metadata Format* used to described the *Referent* is identified by the KEV pair rft_val_fmt=info:ofi/fmt:kev:mtx:pathways. The aip key from the identified *By-Value Metadata Format* conveys the OAIS AIP Identifier. The syntax of the OAIS AIP Identifier itself is a property of the information system.
- Other *Entities* of the *ContextObject* are copied from the initial OAIS DIP bootstrap request.

It should be noted, that in some applications this extra step of interaction with the agent could be by-passed; for instance by allowing the OpenURL *Resolver* to return a specific *ContextObject* instead of a list of *ContextObjects*; the *Resolver* could, for example, do so based on context related information provided by the agent in the initial OAIS DIP bootstrap request.

Once the list of *ContextObjects* has been received by the agent, the agent may choose one and send it back to the OpenURL *Resolver* at baseURL `OpenURL_CIID`. The response to this request is a list of all OAIS DIPs that can be provided for the *Referent*. Again, this list is expressed as an XML container of *ContextObjects*. Each individual *ContextObject* details a specific OAIS DIP request. The XML container is compliant with the aforementioned XML Schema for the XML *ContextObject Format*. For each OAIS DIP format available from the information system, a new *ContextObject* is provided. Each such *ContextObject* has the following characteristics:
- The *Referent* of the *ContextObject* is the OAIS AIP that has been selected by the agent. Again, the *Referent* of the *ContextObject* is described using the combination of an *Identifier Descriptor* and a *By-Value Metadata Descriptor*. The latter conveys the OAIS AIP Identifier of the OAIS AIP being requested by means of the `aip` key. The former expresses the OAIS Content Information Identifier of the OAIS Content Information packaged by that OAIS AIP and is the OAIS Content Information for which the initial OAIS DIP bootstrap service has been requested.
- The *ServiceType* of the *ContextObject* conveys an available OAIS DIP format supported by the information system. The *ServiceType* is provided using an *Identifier Descriptor*. The value of this *Descriptor* is a property of the information system hosting the OAIS AIPs. Though it should be noted that, in order for information systems to transfer content in an interoperable manner, a set of standardized OAIS DIP format – each of which is identified using an *Identifier Descriptor* – will be required. Figure 7 shows the use of an *Identifier Descriptor* to convey a *ServiceType* of the form 'info:pathways/dip.*'.
- Similarly to the OAIS DIP bootstrap request, *Entities* other than *Referent* and *ServiceType* may be provided.

Once the list of *ContextObjects* has been received by the agent, the agent may choose the *ContextObject* of interest and send it back as a service request to the OpenURL *Resolver*.

- ***A specific OAIS DIP requests supported by the information system***: The OpenURL *Resolver* of an information system compliant with Conformance Level 1 of the proposed *OpenURL Application* must support the service requests that it listed in the container of *ContextObjects* in response to the initial OAIS DIP bootstrap service request. Each such service requests results in the response of an OAIS DIP. The syntax of the OAIS DIP is not defined by this *OpenURL Application*.

Conformance Level 1 of the OpenURL *Application* underlying Interface #2 allows for agents to request single sets of content (each of which is packaged in an OAIS DIP) from an information system using OAIS Content Information Identifiers. The strength of this access interface lies in the potential it offers to be deployed across systems and communities. Its use of the ANSI/NISO Standard provides the unique benefit of potentially allowing the tailoring of responses to contextual information conveyed in *ContextObjects*.

### 5.2.2 Conformance Level 2: Requesting Disseminations using NISO OpenURL

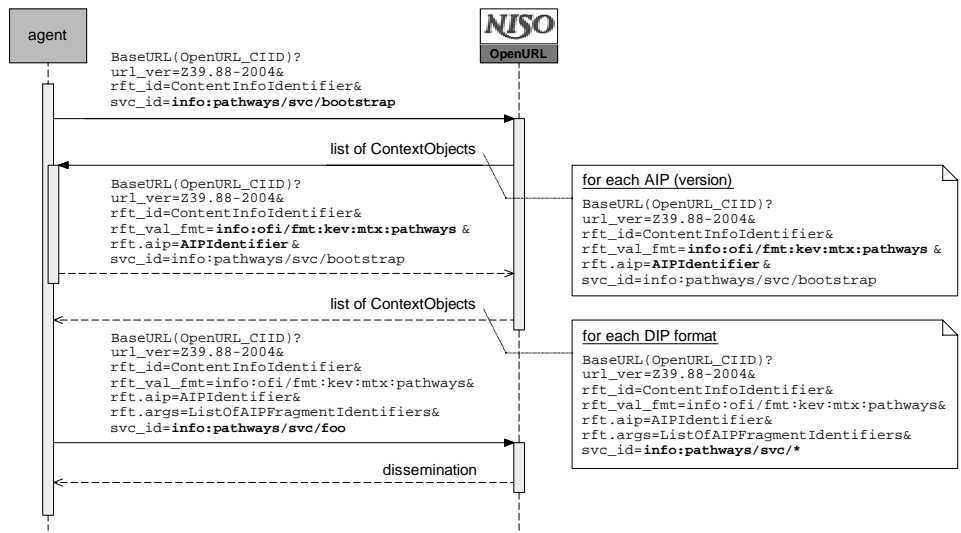

**Fig. 8**. Sequence Diagram of Interface #2, Conformance Level 2:
Requesting Disseminations using OAIS Content Information Identifiers and NISO OpenURL

The Interface of an OpenURL *Resolver* compliant with Conformance Level 2 of this *OpenURL Application* accepts two types of service requests. Similarly to Conformance Level 1, both types of requests are expressed by means of a *ContextObject* that is transported towards the OpenURL *Resolver* at baseURL `OpenURL_CIID`:

- **The interoperable Dissemination bootstrap service request**: The OpenURL *Resolver* of an information system compliant with Conformance Level 2 of this *OpenURL Application* supports the 'Dissemination bootstrap' request. This request is conveyed as a *ContextObject* with the following characteristics:
  - Similarly to Conformance Level 1 of this *OpenURL Application*, the *Referent* of the *ContextObject* is OAIS Content Information stored (as an OAIS AIP) in the information system. The *Referent* is described by means of an *Identifier Descriptor*. Its value is the OAIS Content Information Identifier.

- The *ServiceType* of the *ContextObject* is a service requesting a list of all Dissemination services that can be provided for (files of) the Content Data Object constituting the *Referent*. The *ServiceType* is described by means of an *Identifier Descriptor* with the value 'info:pathways/svc/bootstrap'.
- The *ContextObject* may contain *Entities* other than *Referent* and *ServiceType*, including the *Requester Entity*.

Again, the OAIS Content Information referenced by the *Referent*, may be packaged by multiple OAIS AIPs in the Information system. Therefore, similarly to Conformance Level 1 of this *OpenURL Application*, a separate process is started in which the *OpenURL Application* generates a list of all OAIS AIPs that can be provided for that given OAIS Content Information Identifier, and presents the list to the agent. The list is expressed as an XML container of *ContextObjects*. Each such *ContextObject* has the following charactersitics:
- The *Referent* of the *ContextObject* is an OAIS AIP containing the set of OAIS Content Information for which the initial Dissemination bootstrap service has been requested. The *Referent* is described by the combination of an *Identifier Descriptor* and a *By-Value Metadata Descriptor*. The value of the former is the OAIS Content Information Identifier as has been conveyed by the initial Dissemination bootstrap service. The latter uses the `aip` key to convey the OAIS AIP Identifier of the *Referent*. Again, the *Metadata Format* used to described the *Referent* is identified by the KEV pair `rft_val_fmt=info:ofi/fmt:kev:mtx:pathways`.
- Other *Entities* of the *ContextObject* are copied from the initial Dissemination bootstrap request.

Once the list of *ContextObjects* has been received by the agent, the agent may choose the OAIS AIP of interest and send the corresponding *ContextObject* back to the OpenURL *Resolver* at baseURL `OpenURL_CIID`.

The response to this request, is a list of all dissemination services that can be provided for the (constituents of the) selected OAIS AIP. This list is expressed as an XML container of *ContextObjects* in which each individual *ContextObject* details a specific Dissemination Request. The XML syntax of the container is again expressed by the official XML Schema for the XML *ContextObject Format*. For each Dissemination service available, a new *ContextObject* is provided. Each such *ContextObject* has the following characteristics:
- The *Referent* of the *ContextObject* is a (set of) constituent(s) of the OAIS AIP for which the initial Dissemination bootstrap service has been requested. The *Referent* is described by the combination of two *Descriptors*: 1) an *Identifier Descriptor* conveying the OAIS Content Information Identifier for which the initial Dissemination bootstrap service was requested 2) a *By-Value Metadata Descriptor* consisting of 2 keys. A first key (`aip`) conveys the OAIS AIP Identifier of the

OAIS AIP that has been selected by the agent in the aforementioned process. A second key (`args`) carries one or more Fragment Identifiers pertaining to that OAIS AIP. Each of those Fragment Identifiers points to a constituent (or Content Data Object file) of the OAIS AIP. The syntax of both keys is formally defined by the Metadata Format with identifier `info:ofi/fmt:kev:mtx:pathways`.
- The *ServiceType* of the *ContextObject* conveys an available Dissemination service supported by the information system. The *ServiceType* could be conveyed using an *Identifier Descriptor* or *By-Value* and *By-Reference Metadata Descriptors*. The values of these *Descriptors* is a property of the information system. For example, Figure 8 shows the use of an *Identifier Descriptor* to convey a *ServiceType* of the form '`info:pathways/svc/*`'. In addition, optional *By-Value Metadata Descriptors* could be added that convey arguments for the service.
- The *ContextObject* may contain *Entities* other than *Referent* and *ServiceType*. Again, these *Entities* offer the potential for expressing context related information allowing for the request of context-sensitive Dissemination Requests.

Once the list of *ContextObjects* has been received by the agent, the agent may choose the ContextObject that describes the Dissemination Request of interest and send it back to the OpenURL *Resolver*.

- *A specific Dissemination requests supported by the information system*: The OpenURL *Resolver* of an information system compliant with Conformance Level 2 of the proposed *OpenURL Application* must support the Dissemination Requests that it listed in the container of *ContextObjects* in response to the initial Dissemination bootstrap service request. Each such service requests results in the response of a dissemination of (parts of) a Content Data Object packaged by the referenced OAIS AIP; the result is returned as a MIME-typed stream.

Conformance Level 2 of Interface #2 allows for applications to request dissemination of (parts of) a Content Data Object stored in an information system using OAIS Content Information Identifiers. Again, the merits of this interface are related to its approach that enables cross-community interoperability. In addition, because of its use of the NISO OpenURL Framework, it offers the potential for requesting disseminations that take contextual information into account.

## 6. The access of digital assets in real-life repositories

In current real-life digital repository and archival systems, typically a different access interface is defined per information system. A short overview is provided below.

- *aDORe* [8]: In order to facilitate the retrieval of stored information from the aDORe environment, two access interfaces are introduced.

    First, the *OAI-PMH Federator* provides an OAI-PMH enabled interface through which DIDL documents stored in the aDORe environment can be requested. The identifiers of the DIDL document act as the OAI-PMH *identifiers*. A list of supported packaging formats can be retrieved using a `ListMetadataFormats` request. Currently, the MPEG-21 DIDL and the METS XML-based packaging formats are supported. The documents themselves can be harvested using `GetRecord` and `ListRecords` requests.

    Second, an *OpenURL Resolver* is introduced through which disseminations of constituents of *Digital Item*s can be requested. Requests for disseminatons of datastreams are conveyed using the identifiers of the *Digital Item*s containing the datastreams (represented as *components/resource* constructs) in question. If no *serviceType* is provided, the OpenURL *Resolver* of the aDORe environment will, by default, respond with an XHTML table of contents listing all constituent datastreams of the *Digital Item* as well as the services that are available for them.

- *DSpace* [9,10]: Currently, the only available means for accessing the DSpace system is via its Web user interface. The Web user interface facilitates human-based access by allowing end users to view and submit content and to perform workflow tasks on that content. The Web user interface is implemented using both java Servlets and JSP pages: Java Servlets receive incoming HTTP requests and handle the processing and business logic; and forward the request to a particular JSP for display. Also, at the time of writing, efforts are ongoing at defining a Lightweight Network Interface based on the WebDAV protocol [27] and tailored to the DSpace Data Model. While the DSpace WebDAV interface will allow for the (machine-based) retrieval of digital assets stored in a DSpace repository, it does not seem to provide methods to request versions of digital assets nor disseminations of individual datastreams. The interface also seems to lack the contextual features that can be provided by an interface based on the OpenURL Framework.

- *Fedora* [11,12]: The Fedora system defines two Application Programming Interfaces (APIs) for accessing a Fedora repository: The Fedora-API-A and the Fedora-API-A-LITE. The former is implemented as a SOAP-enabled web service and defines a full blown interface for accessing digital assets stored in the Fedora repository. The access operations include methods to retrieve packagings of Fedora Digital Objects from the Fedora repository, and to discover and request disseminations of datastreams of a Fedora Digital Object. The Fedora-API-A-Lite defines a streamlined version of the Fedora-API-A and is intended to support a REST-like style of access. Both APIs are closely interwoven with the Fedora Data Model.

Based on the OAIS mapping described in Section 3, the two cross-system interface solutions proposed in Section 5 can be implemented for each of the above information systems. A first set of interface is compliant with the solution described in Section 5.1 (Interface #1). This interface uses the OAI-PMH for the Order of OAIS DIPs from an information systems. The specific properties of these interfaces can be summarized as follows:

- The OAI-PMH *identifiers* exposed by each interface are OAIS Content Information Identifiers. As depicted in Figure 2, in the aDORe, DSpace and Fedora information systems, these identifiers correspond with the identifiers of the *Digital Items*, the handles of the DSpace Items and the PIDs of the Fedora Object, respectively.
- The OAI-PMH *metadata format* exposed by each information system corresponds with an OAIS DIP format. The format itself is a property of the information system. Though, it is important to note that in order for information systems to transfer content in an interoperable manner, a standardized and system-agnostic OAIS DIP format is required. The semantics and structure of that format must be known by both the OAI-PMH harvester and the OAI-PMH repository.

A second set of interface is compliant with the solution defined in Section 5.2 (Interface #2). This interface uses the NISO OpenURL framework for the Order of OAIS DIPs (Conformance Level #1) and the request of Dissemination (Conformance Level #2) from an information system. The specific properties of this interface can be summarized as follows:

- The *Referent Identifier Descriptor* corresponds with the OAIS Content Information Identifier of the OAIS Content Information stored in the information system. As shown in Figure 2, in aDORe, DSpace and Fedora, OAIS Content Information Identifiers are mapped to the identifiers of the *Digital Item*, the handle of the DSpace Item and the PID of the Fedora Object, respectively.
- The `aip` key of the *By-Value Metadata Descriptor* describing the *Referent* conveys the OAIS AIP Identifier of an OAIS AIP that packages the OAIS Content Information with the OAIS Content Information Identifier specified by the *Refererent Identifier Descriptor*. As depicted in Figure 2, in aDORe, an OAIS AIP Identifier corresponds with the identifier of a DIDL document; in DSpace, an OAIS AIP Identifier is the combination of the Handle with the file name of the METS document representing the OAIS AIP; and in Fedora, an OAIS AIP can be uniquely identified using the combination of a PID and a local dateTime key.
- The `args` key of the *By-Value Metadata Descriptor* of the *Referent* conveys a set of Fragment Identifiers pertaining to the OAIS AIP (that has been retrieved using the aforementioned `aip` key). In aDORe, DSpace and Fedora, those Fragment Identifiers typically corrrespond with XMLIDs of a DIDL document, a METS document and a FOXML document, respectively.

When paying close attention to the information provided in a *ContextObject* using the above mapping principles, one may notice that in case of both the DSpace and Fedora information systems, redundant information is provided. Indeed, both the *Identifier Descriptor* of the *Referent* and the `aip` key of the *By-Value Metadata Descriptor* of the *Referent* provide a DSpace handle identifier and a Fedora PID, for the DSpace and Fedora system, respectively.

This duplication is merely caused by the fact that versions of content stored in DSpace and Fedora repository systems do not receive identifiers that are unique within those repository systems; but are expressed in function of the OAIS Content Information Identifier of the content being versioned. For example, as described in Section 3 of this paper, in the Fedora repository system, content is versioned through the assignment of a local key that conveys a dateTime of creation or update. This key is unique within the context of the Fedora Digital Object being versioned. The latter is uniquely identified within a Fedora system using its PID. A version of the Fedora Digital Object is uniquely identified within a Fedora system using the combination of the PID and a specific dateTime key.

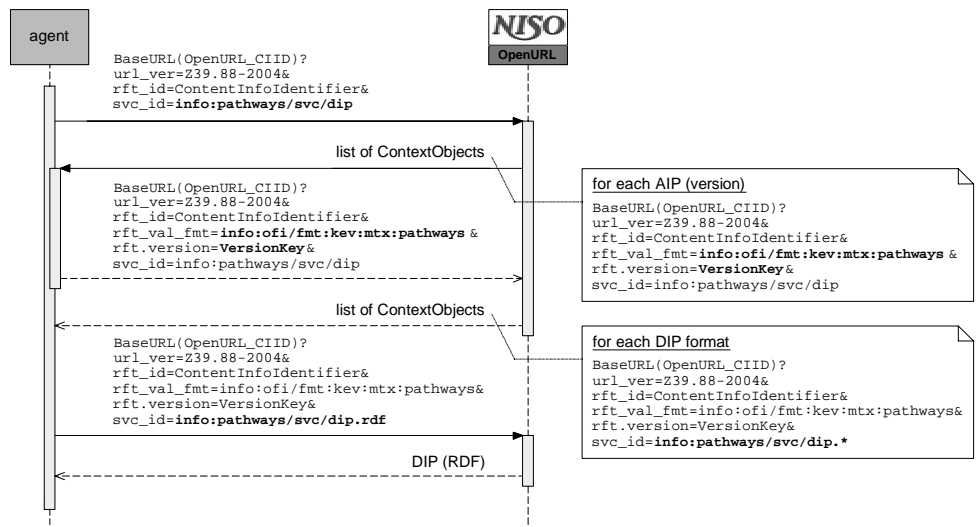

**Fig. 9.** Sequence Diagram of Interface #2, Conformance Level 1 (`version` key): Ordering OAIS DIPs using OAIS Content Information Identifiers and NISO OpenURL

Based on this consideration, a reality-inspired adjustment of Interface #2 can be proposed, by replacing the `aip` key of the *By-Value Metadata Descriptor* of the *Referent* by a `version` key (see Figure 9). The latter allows conveying a repository specific value to distinguish between different *versions* of OAIS Content Information stored in that repository. While the value of the `aip` key, by definition of the OAIS AIP Identifier, must be unique within an information system, the syntax and value of the `version` key is defined as a property of the information system itself. As such, the `version` key could convey a value that is globally unique within the context of an information system (e.g. an OAIS AIP Identifier) or could carry a value that is unique within the context of the OAIS Content Information Identifier (e.g. the Fedora dateTime key).

## 7. Conclusion

This paper has described two formal interfaces that can be deployed across diverse information systems. The first interface is based on the OAI-PMH; the second builds on the NISO OpenURL Framework for Context-sensitive Services. Both interfaces use OAIS Content Information Identifiers, as defined by the OAIS Information Model, as their primary key.

The OAI-PMH interface defines an application-neutral interface for Ordering OAIS DIPs from an archival system. The core characteristics of this interface are:
- The OAI-PMH *identifier* is the OAIS Content Information Identifier of the OAIS Content Information packaged by an OAIS DIP.
- The OAI-PMH *metadata format* is an application neutral, XML-based OAIS DIP format that serializes the requested OAIS DIP. The OAIS DIP format is considered the OAIS Packaging Information of the OAIS DIP.
- The OAI-PMH *datestamp* is the date of creation of the OAIS AIP from which the OAIS DIP is derived.

The use of the OAI-PMH as the protocol for this interface has the following attractive features:
- The ability to selectively harvest batches of OAIS DIPs from archival systems across systems and communities.
- The ease of implementation: OAI-PMH is a lightweight protocol for which several software tools are readily available.
- The ability to augment the OAI-PMH *record* and its associated *metadata* using third party XML Schemas. For example, as described by the authors in [5], W3C XML Signatures can be included in the OAI-PMH responses to facilitate verification of authenticity and integrity of the harvested information. Such capabilities are crucial in a scenario in which trusted mirrors of archives need to be created. Similarly, data providers may associate rights expressions with(in) the returned OAI-PMH *metadata* to indicate how it may be used, shared and modified after it has been harvested. A practical solution on how to convey such rights expressions in described in [28].

The interface based on the NISO OpenURL Framework facilitates responding to two kinds of service requests: First, the Order of individual OAIS DIPs from an information system, and second, the request of Disseminations of datastreams (aka Content Data

Object files). Both levels employ OAIS Content Information Identifiers. The core characteristics of this interface are:
- The *Referent* of the *ContextObject* that bootstraps both requests is a set of OAIS Content Information stored in the information system. It is specified by means of an *Identifier Descriptor* that conveys the identifier of the OAIS Content Information.
- The *ServiceType* of the *ContextObject* conveys the nature of the Dissemination Request. It is specified by means of an *Identifier Descriptor* that is the identifier of the service that is requested as well as by means of an optional *By-Value Metadata Descriptor* that conveys arguments for the service.

The paper also argues that the use of the OpenURL Standard for the specification of an interface for both OAIS DIP Orders and Dissemination Requests has other attractive features:
- The ability to Order OAIS DIPs and request Disseminations of stored content across systems and communities.
- Because the OpenURL Standard is specified in a generic manner, it allows for the same conceptual interfaces to be implemented in different ways as technologies evolve. The concepts underlying the interface remain persistent over time.
- Because the *ContextObject* can contain other *Entities* that are involved in a service request – e.g. *Requester*, *Referrer* and *Resolver* – it offers the potential for requesting context-sensitive Disseminations.